\documentclass{aastex631}

\shorttitle{AASTeX v6.3.1 Sample article}
\shortauthors{Zhang et al.}

\graphicspath{{./}{figures/}}

\begin{document}

\title{Quasiperiodic Super-Alfv\'enic Slippage Along Flare Ribbons Observed by the Interface Region Imaging Spectrograph}

\author[0000-0001-5933-5794]{Yining Zhang}
\affiliation{National Astronomical Observatories, Chinese Academy of Sciences,
Beijing 100101, People's Public of China}
\affiliation{School of Astronomy and Space Science,
University of Chinese Academy of Sciences,
Beijing 100049, People's Public of China}

\author[0000-0001-6655-1743]{Ting Li}
\affiliation{National Astronomical Observatories, Chinese Academy of Sciences,
Beijing 100101, People's Public of China}
\affiliation{School of Astronomy and Space Science,
University of Chinese Academy of Sciences,
Beijing 100049, People's Public of China}
\affiliation{State Key Laboratory of Solar Activity and Space Weather, Beijing 100190,
People's Republic of China}

\author[0000-0002-9534-1638]{Yijun Hou}
\affiliation{National Astronomical Observatories, Chinese Academy of Sciences,
Beijing 100101, People's Public of China}
\affiliation{School of Astronomy and Space Science,
University of Chinese Academy of Sciences,
Beijing 100049, People's Public of China}
\affiliation{State Key Laboratory of Solar Activity and Space Weather, Beijing 100190,
People's Republic of China}

\author[0009-0009-7015-0024]{Xuchun Duan}
\affiliation{National Astronomical Observatories, Chinese Academy of Sciences,
Beijing 100101, People's Public of China}
\affiliation{School of Astronomy and Space Science,
University of Chinese Academy of Sciences,
Beijing 100049, People's Public of China}

\author[0000-0001-5657-7587]{Zheng Sun}
\affiliation{School of Earth and Space Sciences, Peking University, Beijing 100871,
People's Republic of China}
\affiliation{Leibniz Institute for Astrophysics Potsdam, An der Sternwarte 16,
14482 Potsdam, Germany}

\author{Guiping Zhou}
\affiliation{National Astronomical Observatories, Chinese Academy of Sciences,
Beijing 100101, People's Public of China}
\affiliation{School of Astronomy and Space Science,
University of Chinese Academy of Sciences,
Beijing 100049, People's Public of China}
\affiliation{State Key Laboratory of Solar Activity and Space Weather, Beijing 100190,
People's Republic of China}

\correspondingauthor{Ting Li}
\email{liting@nao.cas.cn}

\begin{abstract}
The apparent slipping motion of flare loops is regarded as a key feature of the 3D magnetic reconnection in the solar flares. The slippage with a super-Alfv\'enic speed could be defined as slipping-running reconnection while the slippage with a sub-Alfv\'enic speed is called slipping reconnection. Due to the limitation of the observational instrument temporal resolution, the apparent slippage of the flare loop footpoints along the flare ribbons with super-Alfv\'enic speed is quite rare to our knowledge. In this paper, we report a unique event that exhibits not only the sub-Alfv\'enic slippage, but also the quasiperiodic super-Alfv\'enic slippage of ribbon substructures during a C3.4-class flare (SOL2023-01-18-T15:23), using the high temporal resolution observations of the Interface Region Imaging Spectrograph ($\sim$2 s). The super-Alfv\'enic slippage with a speed of up to $\sim$ 1688 km s$^{-1}$ is directly observed in this study. The calculated period of the apparent super-Alfv\'enic slippage in both ribbons is between 8.4 and 11.9 seconds. This work provides the first observational evidence of the periodicity for the slipping-running magnetic reconnection.

\end{abstract}



\section{Introduction} \label{sec:intro}
Solar flares are eruptive events in the solar system. Up to $10^{33}$ ergs of energy could be ejected into interplanetary space during a time interval of tens of minutes \citep{Aulanier2013}. When magnetic fields of opposite polarities approach each other, the reconfiguration of the magnetic field facilitates the release of magnetic energy. Thus, magnetic reconnection plays a key role in converting the magnetic energy stored by the solar magnetic field into kinetic and thermal energies. The two-dimensional (2D) magnetic reconnection process has been incorporated into the classical CSHKP flare model \citep{Carmichael1964,Sturrock1966,Hirayama1974,Kopp1976}, which has been further verified by observational studies \citep{Ciaravella2008,Cheng2018,Tan2020}.

It should be noted, however, that 2D models can be considered as a simplified model for magnetic reconnection, which is intrinsically a 3D process \citep{Priest1995,Demoulin19961}. Under 3D conditions, the change in magnetic connectivity is continuous at the quasi-separatrix layer (QSL), where the magnetic field has an extremely high gradient. The magnetic lines undergo a series of reconnections with the footpoints slipping at the photosphere within the QSL structure \citep{Priest1995,Demoulin19962}. This causes the apparent slipping motion of the flare loop footpoints in the flare ribbons. Thus, the slipping motion is considered as the key feature for 3D magnetic reconnection model different from 2D circumstances. \citet{Aulanier2006} performed magnetohydrodynamic (MHD) simulations and found the continuous slipping of magnetic field lines along the current layer of QSL with a super-Alfv\'enic speed. This process is called slipping-running reconnection. As for the slippage with a speed lower than the local Alfv\'enic speed, it is called slipping reconnection. \citet{Janvier2013} performed 0-$\beta$ MHD simulations and showed that the slipping-running reconnection happens while one of the footpoints of the magnetic field line is fixed at the strong current region and the other one slips along the QSL, which is further validated by the observations in \citet{Janvier2014}.

With the development of the observational instruments, the apparent slipping motion has been verified by more and more studies. \citet{Fletcher2004} tracked the motion of the flare footpoints in the chromosphere with an average speed of 15 km s$^{-1}$ using the Transition Region and Coronal Explorer satellite at 1600 \AA~waveband. \citet{Dudik2014} identified the slipping reconnection occurring during an eruptive X-class flare using Solar Dynamics Observatory (SDO, \citealt{Pesnell2012})/Atmospheric Imaging Assembly (AIA, \citealt{Lemen2012}) data. The maximum slipping speed in their study is 136 km s$^{-1}$ at 131 \AA~waveband. \citet{Jing2017} observed the apparent slipping motion of coronal loops along the flare ribbon during an M-class flare. The slipping speed is calculated to be 4-217 km s$^{-1}$ from AIA 211 \AA~images. \citet{Lorincik2019} reported the slipping motion of the flare ribbon kernels at AIA 1600 and 304 \AA~wavebands, which are possibly related to the flare loop footpoints with a higher velocity of up to 450 km s$^{-1}$. As for a special type of circular flares, \citet{Li2018} and \citet{Huang2024} both studied the slipping motion of the bright substructures which could be regarded as the flare loop footpoints in the circular ribbon and found a speed of about 40 km s$^{-1}$ to 220 km s$^{-1}$. Due to the instrumental limitations, especially the temporal resolution, only the slipping motion with sub-Alfv\'enic speed has been reported until this year. Recently, \citet{Lorincik2025} reported the slipping speed of thousands of kilometers per second by using the Interface Region Imaging Spectrograph (IRIS, \citealt{Depontieu2014}) high temporal cadence data ($\sim$2 s). They provide the first observational evidence for slipping-running magnetic reconnection, showing the super-Alfv\'enic slipping motion of ribbon substructures.

Quasiperiodic pulsations (QPPs) are another commonly observed feature in solar flares that show quasiperiodic oscillations in the light curves. QPPs are characterized by their distinct periodicities, which can range from sub-seconds to several minutes, and are widely regarded as the probes for revealing the physical processes of energy release and particle acceleration during flares \citep{Nakariakov2009,McLaughlin2018,Zimovets2021}. Interestingly, some slipping motions along the flare ribbons also show a quasiperiodic pattern according to previous studies. \citet{Li2015} first identified the quasiperiodic slipping motion along the flare ribbons using SDO/AIA and IRIS. The calculated slippage speed is in the range of 20-110 km s$^{-1}$ with a period of 3-6 minutes. \citet{Li2016} calculated the slipping speed during an eruptive X-class flare to be 30-40 km s$^{-1}$ using IRIS 1400 \AA~and 1330 \AA~data with a period of about 130 s. \citet{Jeffrey2018} gave the first report of rapid oscillatory behavior (period $\sim$ 10 s) in the line broadening of Si \uppercase\expandafter{\romannumeral4} line in the lower atmosphere using high-cadence IRIS data, which is possibly related to the interacting waves (turbulent velocity fluctuations). This work provides a new sight that the turbulence could produce QPPs in the flaring region. Recently, \citet{Zhang2024} studied a C-class flare and found the simultaneous occurrence of the quasiperiodic slipping motion of the ribbon substructures and the intensity oscillations of the flare loops with a common period of 4-5 minutes.

However, it remains unclear whether periodicity exists in the rare super-Alfv\'enic slipping motion events. Here, we report a unique event that the super-Alfv\'enic slipping motion of ribbon substructures displays the quasiperiodic pattern with a period of about 10s, as well as the simultaneous observation of sub-Alfv\'enic slippage in the flare ribbons. This paper is organized as follows: In Section \ref{sec:obs} we describe the observational results and the associated analysis of the periodicity of this event. In Section \ref{sec:dis} we discuss the potential mechanisms responsible for the super-Alfv\'enic slipping motion in the ribbons, the quasiperiodic behaviors of the slipping-running magnetic reconnection, and the related velocity error analysis. Finally, Section \ref{sec:con} gives the conclusion of this work.

\section{Observational Results} \label{sec:obs}
\begin{figure*}[!ht]
    \centering
    \includegraphics[width=0.90\textwidth]{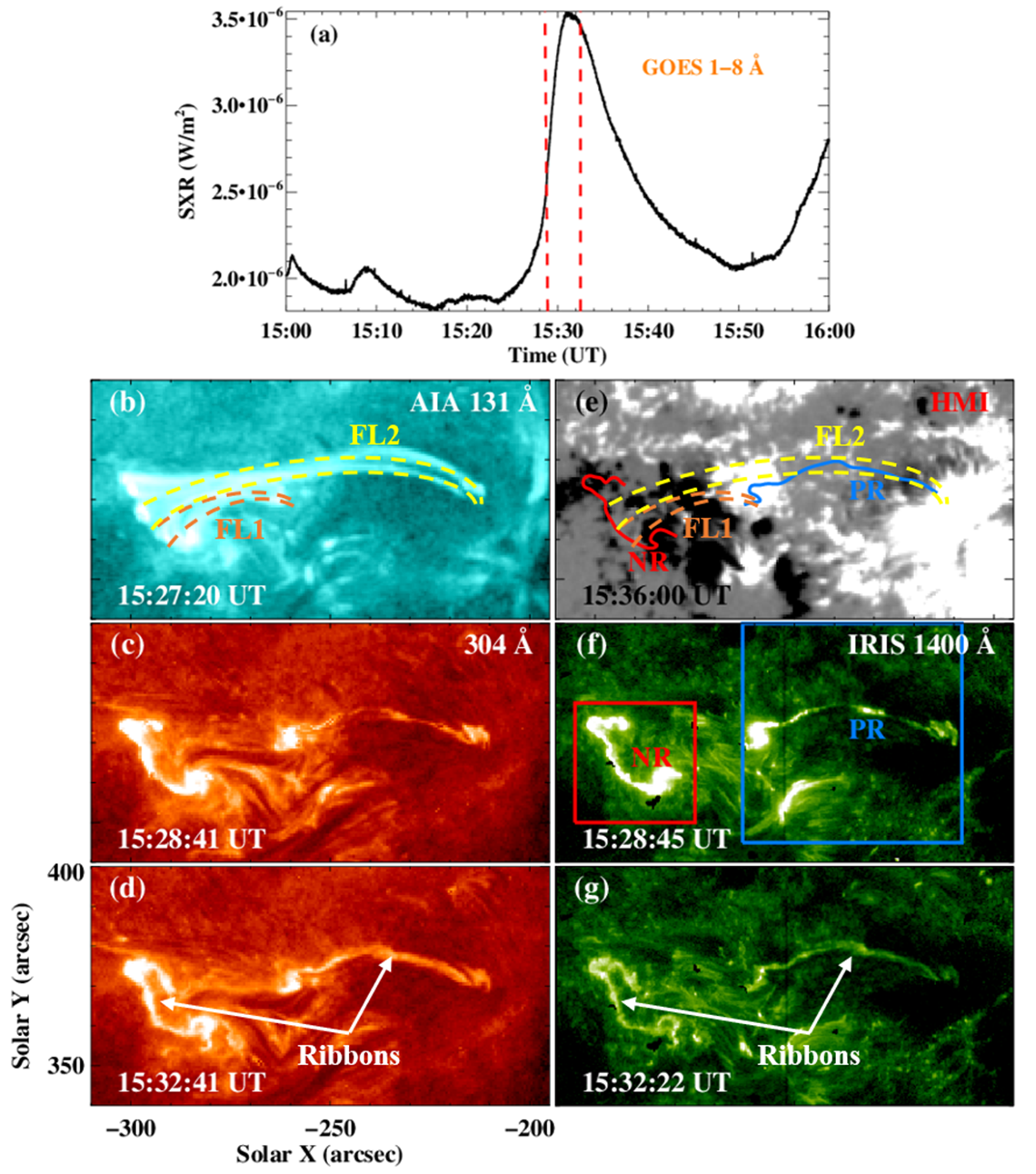}
    \caption{Overview of the C3.4-class flare SOL2023-01-18-T15:23. Panel (a): GOES SXR data at 1-8 \AA~waveband from 15:00 to 16:00 UT. Red dashed lines indicate the time of interest in panels (f) and (g), respectively. Panel (b): SDO/AIA 131 \AA~image with a field of view (FOV) of 105"$\times$60". Panels (c)-(g): SDO/AIA 304 \AA~images at different time, HMI line-of-sight magnetogram and IRIS 1400 \AA~images with the same FOV as in panel (b). Orange and yellow dashed curves in panels (b) and (e) denote two groups of flare loops `FL1' and `FL2', respectively. The red and blue solid curves in panel (e) denote the negative ribbon (NR) and positive ribbon (PR) of the flare, respectively. Red and blue squares in panel (f) denote the FOV of panels (a)-(d) and panels (e)-(h) in Figure \ref{fig:fig2}, respectively. The animation of this figure includes HMI magnetogram, AIA 131 \AA, 171 \AA~and 1600 \AA~images from 15:15 to 15:45 with a video duration of 6 s.}
    \label{fig:fig1}
\end{figure*}

The flare event is selected based on the following criteria. First, the IRIS slit-jaw images (SJIs) centered at 1400 \AA~show clear observations of flare ribbons with a time cadence $\leq$2 s. Second, the ribbon substructures exhibit the apparent slipping motions along the flare ribbons. Third, the slippage of the ribbon substructures shows the clearly discerned quasiperiodic pattern. Based on the above three criteria, the event SOL2023-01-18T15:23 is selected to study the quasiperiodic Super-Alfvénic slipping motion in this work.
On 2023 January 18, a C3.4-class flare occurred in the domain of NOAA active region (AR) 13192. As indicated by the GOES SXR data (see Figure \ref{fig:fig1}(a)), the flare initiated at 15:23 UT, peaked at 15:31 UT and ended at 15:38 UT. This flare showed a typical two-ribbon configuration (see Figures \ref{fig:fig1}(b)-(g) and the associated animation). In high temperature wavebands such as AIA 94 \AA~(log$T$[K]$\sim$6.85) and 131 \AA~(log$T$[K]$\sim$7.05, \citealt{O'Dwyer2010}), two groups of flare loops could be clearly seen with a shorter group `FL1' and a longer group `FL2' (see Figures \ref{fig:fig1}(b) and (e)). As shown in the HMI magnetogram (see Figure \ref{fig:fig1}(e)), the two ribbons consisted of a zip-shaped eastern ribbon located in the negative magnetic polarity region (named NR) and a western long ribbon located in the positive polarity region (named PR). 

\begin{figure*}[!ht]
    \centering
    \includegraphics[width=0.90\textwidth]{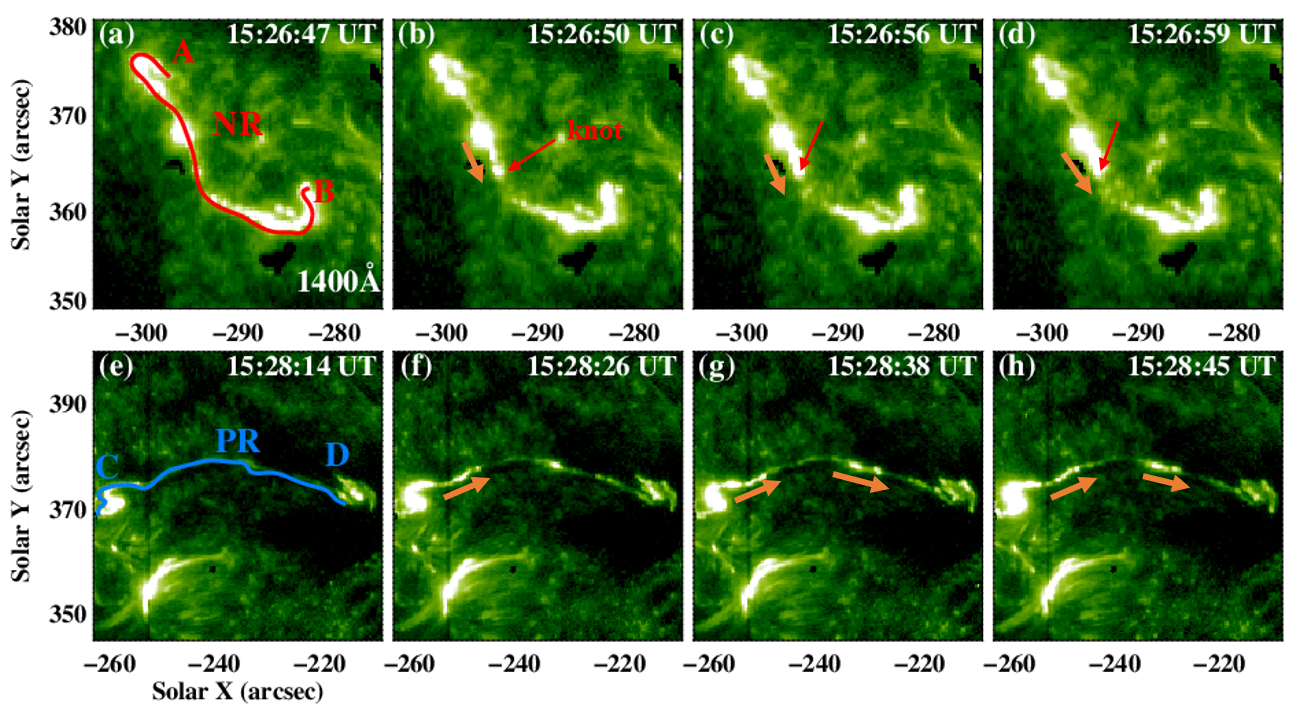}
    \caption{Slipping motion of the ribbon substructures at IRIS 1400 \AA~images. Panels (a)-(d): Fine substructures in the NR region. Panels (e)-(h): Fine substructures in the PR region. Red and blue curves with `A-B' and `C-D' in panels (a) and (e) denote the slits used to obtain the stack plots in Figures \ref{fig:fig3}(c)-(d) and (e)-(f), respectively. Orange arrows in panels (b)-(d) and (f)-(h) describe the slipping direction in the ribbons. The animations of this figure include IRIS 1400 \AA~images at NR and PR regions from 15:26 to 15:34 UT, respectively. The duration of both videos is 30 s.}
    \label{fig:fig2}
\end{figure*}

In this study, we use high temporal resolution data from IRIS with a typical cadence of 2 s. The slipping motion of the ribbon substructures along both PR and NR started at 15:26 UT just after the initiation of the flare (see Figure \ref{fig:fig2} and the associated animations of PR and NR). In Figures \ref{fig:fig2}(a)-(d), bright knots appear in the ribbon NR with a slipping direction from the northeastern to the southwestern at a speed of about 100-200 km s$^{-1}$. However, we can also observe the fast moving brightening propagation in the NR which could not be located or tracked directly. Simultaneously, there are also many fast moving brightening propagations in the PR (see Figures \ref{fig:fig2}(f)-(h)). These fine substructures in flare ribbons are generally regarded as the footpoints of the magnetic field lines or flare loops \citep{Fletcher2004, Dudik2014, Li2015, Jing2017, Li2018, Lorincik2019}.

\begin{figure*}
    \centering
    \includegraphics[width=0.90\textwidth]{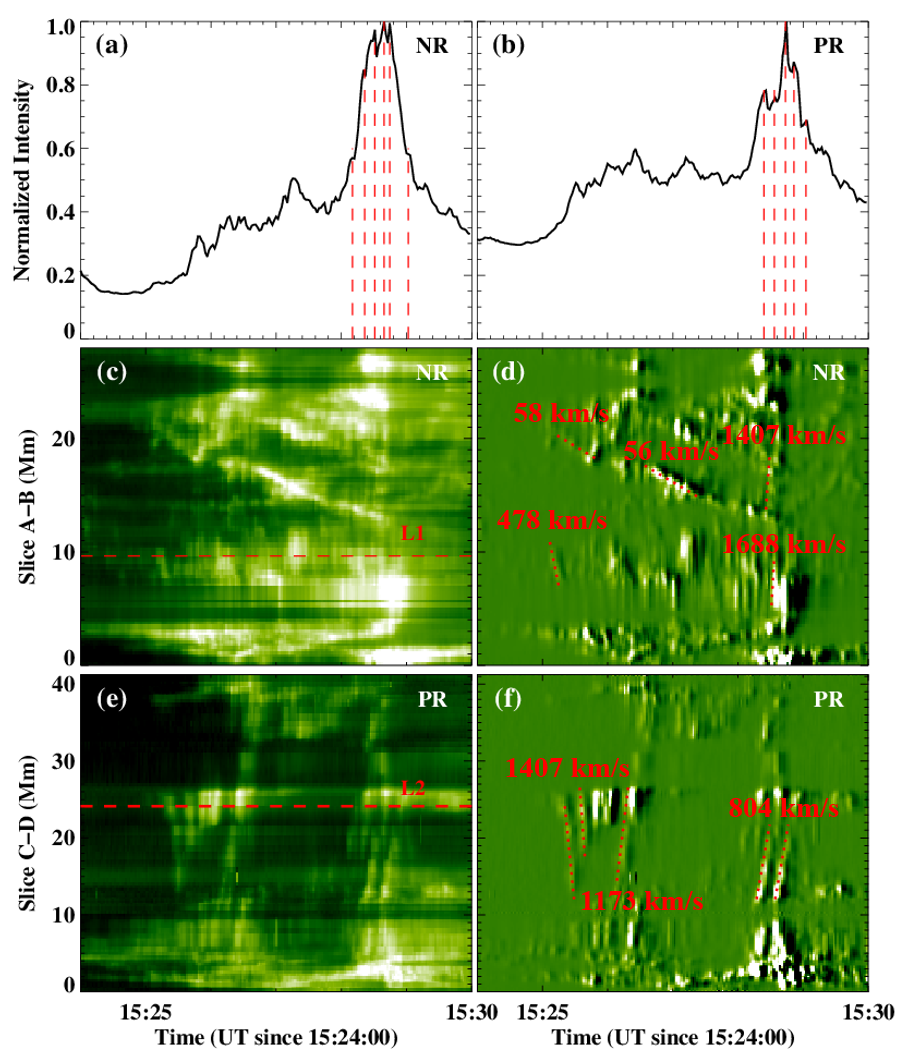}
    \caption{IRIS 1400 \AA~time–distance stack plots of the apparent slipping motion along the ribbons. Panels (a)-(b): The normalized integral intensity of the FOV in Figures \ref{fig:fig2}(a)-(d) and (e)-(h) from 15:24 to 15:30 UT, respectively. Red dashed lines mark the oscillation peaks in the light curves. Panels (c)-(d): Stack plots along slice `A-B' of the original images and running difference images, respectively. Panels (e)-(f): Stack plots along slice `C-D' of original images and running difference images, respectively. Red dashed lines in panels (c) and (e) denote the positions to obtain the light curves `L1' in Figures \ref{fig:fig5}(a) and `L2' in \ref{fig:fig5}(d).}
    \label{fig:fig3}
\end{figure*}

The normalized integral intensities of the FOV in Figures \ref{fig:fig2}(a)-(d) and (e)-(h) are shown in Figures \ref{fig:fig3}(a) and (b) from 15:24 to 15:30 UT, respectively. It is noteworthy that both light curves show clear oscillations with an overall ascending trend in both the PR and NR regions. The light curves show 6 peaks in NR and 5 peaks in PR during 15:28 to 15:29, which give a corresponding period of about 10 s. To study the temporal evolution of these slipping substructures in both ribbons, we draw the slices `A-B' and `C-D' to obtain the time-distance plots in Figure \ref{fig:fig3}. Several tens of bright stripes can be clearly seen from the  time-distance plots, representing the moving substructures in the ribbons. The typical movement distance is more than 5 Mm from the stack plots. The bidirectional slipping is represented by the different oblique direction of the bright stripes. The stripes representing the fast and slow slippage are presented simultaneously in Figure \ref{fig:fig3}. We calculate the associated velocity of each stripe from the both the original and the running-difference (difference is 1 frame) stack plots. The speed of the fast slippage ranges from $\sim$478 to 1688 km s$^{-1}$ and the slow slippage has a speed of about 50 km s$^{-1}$.

\begin{figure*}
    \centering
    \includegraphics[width=0.95\textwidth]{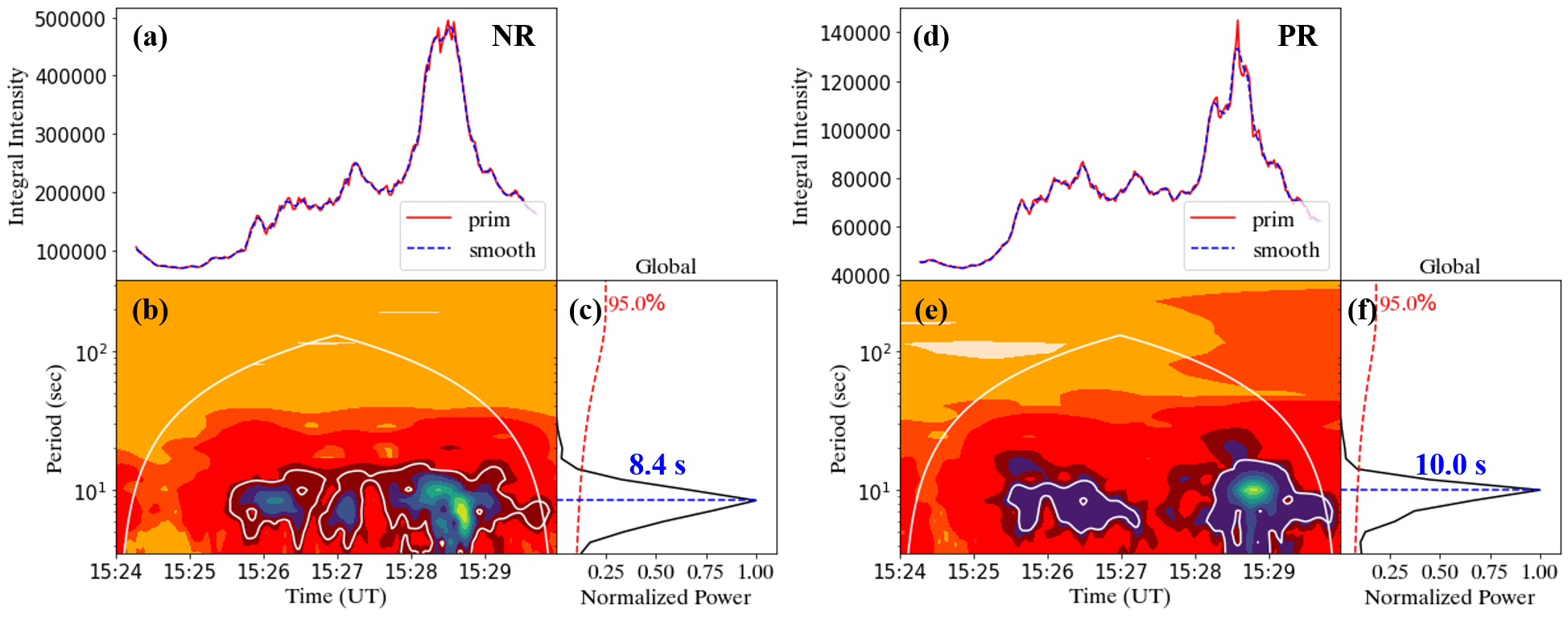}
    \caption{Wavelet analysis of the integral intensity in the NR and PR regions. Panel (a): Red solid and blue dashed lines indicate the original and smoothed light curves of the `NR' region in Figure \ref{fig:fig3}(a), respectively. Panel (b): The wavelet map of the detrended intensity calculated from the difference between the two light curves in panel (a) from 15:24 to 15:30 UT. Panel (c): The corresponding global power. The red dashed line represents the 95\% confidence level, and the blue dashed line represents the period above the 95\% confidence level with the maximum power (8.4 s). Panels (d)-(f): Similar analysis to panels (a)-(c), but for the PR region.}
    \label{fig:fig4}
\end{figure*}

Figure \ref{fig:fig4} shows a wavelet analysis of the integral intensity in the NR and PR regions (see Figures \ref{fig:fig3}(a) and (b)) based on the Morlet transform using Python \citep{Torrence1998}. Both regions show oscillatory behaviors in the light curves (see red lines in Figures \ref{fig:fig4}(a)\&(d)). To enhance the oscillations in the light curves, we use the detrended light curves which are calculated from the difference between the original light curves (red solid lines in Figures \ref{fig:fig4}(a)\&(d)) and the smoothed light curves (blue dashed lines in Figures \ref{fig:fig4}(a)\&(d)), for the wavelet analysis \citep{Kupriyanova2010, Kupriyanova2013, Gruber2011, Auchere2016}. The average temporal cadence $\Delta $t is about 1.71 seconds in the interval 15:24-15:30UT. The smoothing window is set to 10 points in our detrending process, which gives a window length of 15.4 s (similar smoothing process to \citealp{Inglis2008,Nakariakov2010,Lidong2015}). As a result of the global computation, the light curve in the NR region shows a maximum period of 8.4 s, which exceeds the 95\% confidence level (see Figure \ref{fig:fig4}(c)). Similarly, the PR region shows oscillations with a period of about 10.0 s (see Figure \ref{fig:fig4}(f)). Note that the oscillation signals occurred mainly during the time interval 15:25-15:30 UT, which is at the impulsive phase of the flare (see Figure \ref{fig:fig1}(a)).

\begin{figure*}
    \centering
    \includegraphics[width=0.95\textwidth]{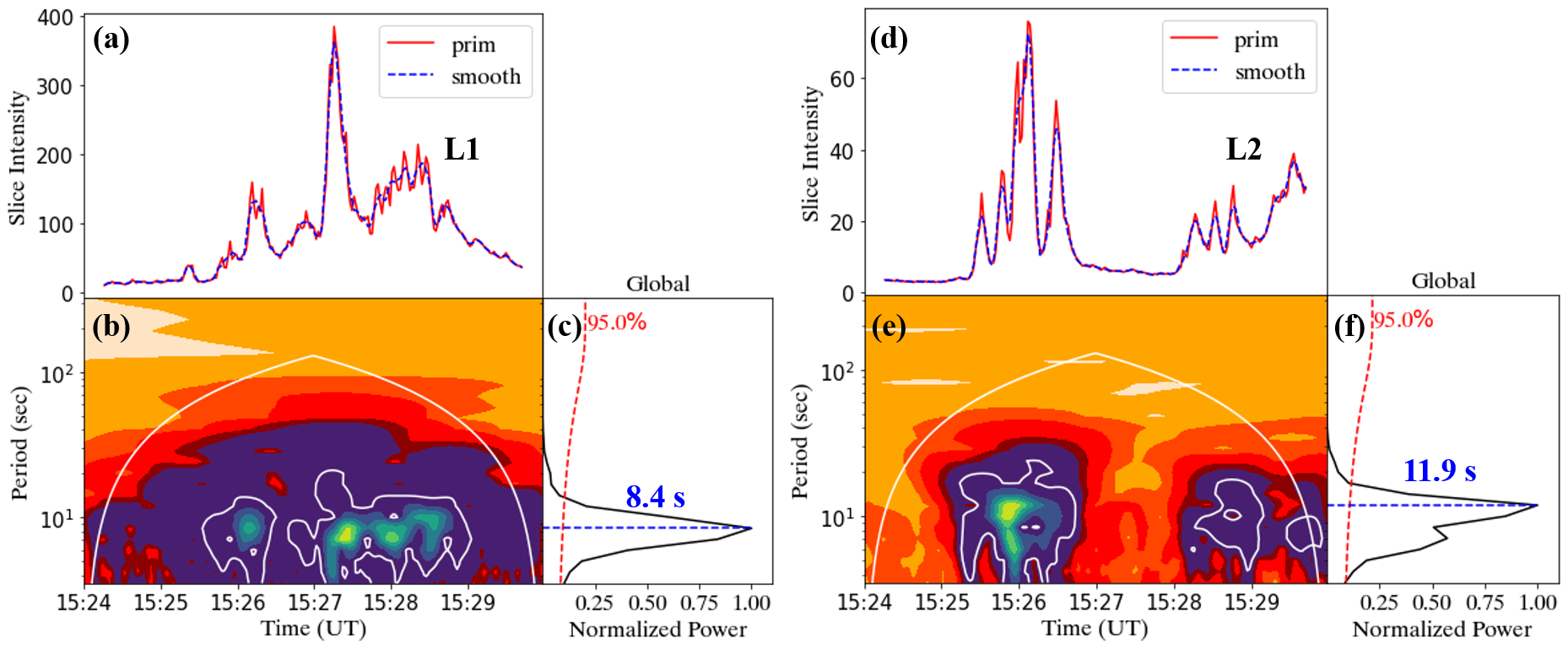}
    \caption{Wavelet analysis of the fast slipping motion at certain positions along the ribbons NR and PR. Similar to Figure \ref{fig:fig4}, but for the light curves `L1' and `L2'.}
    \label{fig:fig5}
\end{figure*}

To further study the oscillatory behavior at a certain position, we select two positions from both NR and PR regions (see red dashed lines in Figures \ref{fig:fig3}(c) and (e)) and calculate their light curves of `L1' and `L2' (see the red solid lines in Figures \ref{fig:fig5}(a) and (d)) in the original IRIS 1400 \AA~images with the same time interval as in Figure \ref{fig:fig4}. The criterion for selection is the position that contains as many bright stripes as possible. Both original light curves underwent apparent oscillations from about 15:26 UT. With several peaks and troughs, the light curve `L1' entered the decay phase at about 15:29 UT. The Morlet spectrum is shown in Figure \ref{fig:fig5}(b), where the white line is the 95\% confidence level of the spectrum. As a result of the global power map calculation, the blue dashed line indicates the epoch of maximum power, corresponding to an oscillation period of about 8.4 s (see Figure \ref{fig:fig5}(c)). Note that its power exceeds the 95\% confidence level. Different from the general trend of `L1', `L2' reached its peak at about 15:26 UT, and the oscillations occurred mainly between 15:25-15:26 UT and 15:28-15:30 UT. We perform the similar wavelet analysis for the light curve `L2' and obtain a period of 11.9 s (see Figure \ref{fig:fig5}(f)). The period of `L2' also exceeds the 95\% confidence level.

\section{Discussion}   \label{sec:dis}

Magnetic reconnection process plays a key role in the energy release in solar flares. Generally, the speed of the reconnection inflows in solar flares is $u_i \sim 10^{-3}-0.05$~$v_A$, where $v_A$ is the Alfv\'enic speed \citep{Aulanier2006}, while the outflow speed can be regarded as the local Alfv\'enic speed \citep{Yokoyama2001,Ni2020}. Previous studies gave the typical values of $v_A$ of 1000 km s$^{-1}$ in the low coronal environment \citep{Hollweg1978,Russell2013}. In our study, the calculated slipping velocity of the ribbon bright knots exceeds 1000 km s$^{-1}$, which can be classified as super-Alfv\'enic slipping. \citet{Lorincik2025} suggested that only the temporal resolution higher than 4 s can capture the flare substructures with a velocity higher than 1000 km s$^{-1}$. Thus, it is only by using these high temporal resolution data from the IRIS observations that we can distinguish the super-Alfv\'enic slipping of the substructures in the flare ribbons. The parameters that determine the apparent slipping velocity remain unclear till now. \citet{Aulanier2006} first discussed that the apparent slipping velocity should be comparable to $Q^{1/2}$, where $Q$ represents the squashing degree of the QSL \citep{Titov2002}. \citet{Janvier2013} found that the apparent slipping speed of the field lines is proportional to the QSL mapping norm $N$ of the QSL \citep{Priest1995}. In our study, ribbon NR is located at the border of two different magnetic domains (`FL1' and `FL2' in Figure \ref{fig:fig1}(a)) with a high magnetic field gradient. This can lead to the super-Alfv\'enic slippage in the NR. Another factor that can determine the apparent slipping speed is the magnetic field strength. \citet{Lorincik2019} found a rough anticorrelation between the apparent slipping speed and the magnetic field strength. They found only the slowest moving flare kernals through the strong-B region in the 131 \AA~observations. Our observations for the apparent super-Alfv\'enic slippage in the ribbon PR are consistent with this conclusion. We observe the super-Alfv\'enic slippage mainly in the central part of slice `C-D' (see Figures \ref{fig:fig3}(e)\&(f)) and not in the tips of the PR. We can see that the two tips of PR are located in the sunspot region, which is generally associated with a stronger magnetic field $B$ in the photosphere. Thus, the super-Alfv\'enic slippage appears only in the middle part of the stack plots with weak magnetic field.

The integral intensity light curves of both ribbons (Figures \ref{fig:fig3}(a)\&(b)) and the light curves at specific positions in the stack plots (Figures \ref{fig:fig5}(a)\&(d)) both show the similar quasiperiodic oscillations with periods close to 10 s. However, our calculated period is much shorter than those in previous studies of the quasiperiodic phenomena in the flare ribbons \citep{Li2015,Brannon2015,Zhang2024} with periods of several minutes. We suggest that this can be explained by two aspects. On the one hand, the high temporal resolution of IRIS could capture the short-period component of the slipping motion along the flare ribbons, which could not be captured by previous observational instruments with lower temporal resolution. On the other hand, the apparent super-Alfv\'enic slippage with higher speeds could have a quasiperiodic slipping nature with a much shorter period. Since the bright dot-like structures generally correspond to the footpoints of the flare loops \citep{Dudik2014,Dudik2016,Lorincik2019}, this quasiperiodic slippage could be attributed to the intermittent slipping-running magnetic reconnection process.

Another point needs to be considered is the method we use to calculate the slipping velocity from the stack plots along the ribbons (see Figure \ref{fig:fig3}). This approach can introduce extra error in the identification and fitting of the bright stripes in the stack plots. For instance, we calculate the slipping speed by:
\begin{equation}
    v_{slip} = \frac{x_2-x_1}{t_2-t_1}
    \end{equation}
where $x_1$, $x_2$, $t_2$ and $t_1$ are the start and end distances along the slice and the start and end times of the stripes, respectively. This can be used to calculate the slip velocity deviation:
\begin{equation}
    \Delta v_{slip} = \sqrt{(\frac{\Delta x_1}{t_2-t_1})^2 + (\frac{\Delta x_2}{t_2-t_1})^2 + (\frac{(x_2-x_1)\Delta t_1}{(t_2-t_1)^2})^2
    + (\frac{(x_2-x_1)\Delta t_2}{(t_2-t_1)^2})^2} 
\end{equation}
where $\Delta x_1$, $\Delta x_2$, $\Delta t_1$ and $\Delta t_2$ are the error of the start and end positions and the start and end times of the fitted stripe, respectively. Note that for a given observing instrument, the errors in distance and time are fixed. Thus, the maximum velocity is accompanied by a maximum velocity error. For the maximum speed calculated from Figure \ref{fig:fig3}, which is 1688 km s$^{-1}$, we take the position error to be half of the spatial resolution of IRIS and the time error to be half of the temporal resolution of the IRIS data. If we set $\Delta x_1=0.1664$ arcsec, $\Delta x_2=0.1664$ arcsec, $\Delta t_1=1$ s and $\Delta t_2=1$ s, the associated maximum speed error is $\Delta v_{slip}=799$ km s$^{-1}$. Then the maximum slipping speed in our calculation is $1688\pm799$ km s$^{-1}$, which means that this estimation method should be further optimized for a better accuracy in obtaining the slipping speed. For the stripes with a higher moving speed, their shape in the stack plots tends to become a vertical stripe, which makes it more difficult to determine the time interval of the stripe. Since the slipping time only takes 1-2 time pixels, it is quite difficult to determine the exact start and end times. We also calculate the speed errors for the longer stripes in Figure \ref{fig:fig3} with speeds of 1173 km s$^{-1}$ and 804 km s$^{-1}$ which have trajectories on the orders of 10". The related speeds with error are $1173\pm139$ km s$^{-1}$ and $804\pm82$ km s$^{-1}$, respectively. For this consideration, higher temporal resolution observations are needed to capture the super-Alfv\'enic slippage in the flare ribbons in the future.

\section{Conclusion}   \label{sec:con}
In this paper, we use data from IRIS and SDO to observe the apparent slipping motion that occurs during the C3.4-class flare event SOL2023-01-18T15:23. From the IRIS high temporal cadence data, we identify the slipping motion in the two flare ribbons NR and PR. By calculating the apparent slipping velocity of the ribbon substructures, we obtain both the sub-Alfv\'enic slippage with speeds of 50 to 400 km s$^{-1}$ and the super-Alfv\'enic slippage which exceeds 1000 km s$^{-1}$ in IRIS 1400 \AA~images. The classification of \citet{Aulanier2006} suggests that when the apparent slipping speed exceeds the local Alfv\'enic speed, the slipping should be called slipping-running magnetic reconnection. In addition, the bright dot-like substructures slip quasiperiodically with a period of about 10 s. This could be explained by the intermittent slipping-running magnetic reconnection. To our knowledge, this quasiperiodic behaviour of the slipping-running magnetic reconnection process is reported for the first time. In the future, we will carry out statistic studies on more 
super Alfv\'enic slipping motion events based on high temporal resolution data, in order to explore the spatiotemporal characteristics and the statistical distribution of their periodic features.

\section*{Acknowledgments}

We thank the anonymous referee for valuable comments and suggestions to improve the quality of this paper. We acknowledge the IRIS which is a NASA small explorer mission developed and operated by LMSAL with mission operations executed at NASA Ames Research center and major contributions to downlink communications funded by the Norwegian Space Center (NSC, Norway) through an ESA PRODEX contract. We also acknowledge SDO/AIA and HMI for providing data. This work is supported by the National Key R\&D Programs of China (2022YFF0503800), the National Natural Science Foundations of China (12222306, 12273060), B-type Strategic Priority Program of the Chinese Academy of Sciences (Grant No.XDB0560000) and the Youth Innovation Promotion Association of CAS (2023063).

\bibliography{sample631}{}
\bibliographystyle{aasjournal}

\end{document}